\documentclass[twocolumn]{article}

\usepackage[a4paper, margin = 0.5in ]{geometry}
\usepackage{graphicx}
\usepackage{dblfloatfix} 
\usepackage[dvipsnames]{xcolor}
\definecolor{mygray}{gray}{0.90}

\usepackage{lipsum}

\title{{\bf Magnetization Transfer in Magnetic Resonance Fingerprinting}}

\author{Tom Hilbert$^{1,2,3,*}$, Ding Xia$^{4,5}$, Kai Tobias Block$^{4,5,6}$, Zidan Yu$^{4,5,7}$, Riccardo Lattanzi$^{4,5,7}$, \\Daniel K. Sodickson$^{4,5,7}$, Tobias Kober $^{1,2,3}$, Martijn~A.~Cloos$^{4,5,7}$}

\date{}

\begin{document}

\twocolumn[
  \begin{@twocolumnfalse}
  
  {\centering PREPRINT - SUBMITTED TO MAGNETIC RESONANCE IN MEDICINE}\\
%  \vspace{-0.5cm}
    \maketitle

\vspace{-0.5cm}
{\tiny $^1$ Advanced Clinical Imaging Technology, Siemens Healthcare AG, Lausanne, Switzerland.}\\
{\tiny $^2$ Department of Radiology, Lausanne University Hospital and University of Lausanne, Lausanne, Switzerland.}\\
{\tiny $^3$ LTS5, \'Ecole Polytechnique F\'ed\'erale de Lausanne, Lausanne, Switzerland.}\\
{\tiny $^4$ Bernard and Irene Schwartz Center for Biomedical Imaging, Department of Radiology, New York University School of Medicine, New York, NY, USA.}\\
{\tiny $^5$ Center for Advanced Imaging Innovation and Research (CAI2R), Department of Radiology, New York University School of Medicine, New York, NY, USA.}\\
{\tiny $^6$ Department of Radiology, University Hospital Basel, Basel, Switzerland.}\\
{\tiny $^7$ The Sackler Institute at the New York University School of Medicine, 550 First Avenue, New York, NY 10016, USA.}

\vspace{0.5cm}
    \centering
    \begin{minipage}{.25\textwidth} 
{\bf \small Correspondence}\\
{\tiny Tom Hilbert\\
Siemens Healthineers,\\
EPFL Innovation Park QI-E, 1015 Lausanne, Switzerland.\\
Email: \\tom.hilbert@siemens-healthineers.com\\
Twitter: @TomHilbertMRI\\}
\vspace{0.5cm}\\
{\bf \small Funding information}\\
{\tiny This work was supported by funding from the National Institutes of Health ( R21~EB020096, R01~AR070297,  R01~EB026456,  P41~EB017183 ).}
    \end{minipage}%
    \begin{minipage}{.05\textwidth}
    ~
    \end{minipage}%        
{\colorbox{mygray}{ \begin{minipage}{0.67\textwidth}
{\bf Purpose:} To study the effects of magnetization transfer (MT, in which a semi-solid spin pool interacts with the free pool), in the context of magnetic resonance fingerprinting (MRF).\vspace{0.2cm}\\ 
{\bf Methods:} Simulations and phantom experiments were performed to study the impact of MT on the MRF signal and its potential influence on $T_1$ and $T_2$ estimation. Subsequently, an MRF sequence implementing off-resonance MT pulses and a dictionary with an MT dimension by incorporating a two-pool model were used to estimate the fractional pool size in addition to the $B_1^+$, $T_1$, and $T_2$ values. The proposed method was evaluated in the human brain.\vspace{0.2cm}\\ 
{\bf Results:} Simulations and phantom experiments showed that an MRF signal obtained from a cross-linked bovine serum sample is influenced by MT. Using a dictionary based on an MT model, a better match between simulations and acquired MR signals can be obtained (NRMSE 1.3\% versus 4.7\%). Adding off-resonance MT pulses can improve the differentiation of MT from $T_1$ and $T_2$. In-vivo results showed that MT affects the MRF signals from white matter (fractional pool-size $\approx$16\%) and gray matter (fractional pool-size $\approx$10\%). Furthermore, longer $T_1$ ($\approx$1060 ms versus $\approx$860 ms) and $T_2$ values ($\approx$47 ms versus $\approx$35 ms) can be observed in white matter if MT is accounted for.\vspace{0.2cm}\\ 
{\bf Conclusion:} Our experiments demonstrated a potential influence of MT on the quantification of $T_1$ and $T_2$ with MRF. A model that encompasses MT effects can improve the accuracy of estimated relaxation parameters and allows quantification of the fractional pool size.
    \end{minipage}
    }}
\\ \vspace{0.5cm}
  \end{@twocolumnfalse}
]

%\maketitle

%\footnotetext[1]{\it Advanced Clinical Imaging Technology, Siemens Healthcare AG, Lausanne, Switzerland.}
%\footnotetext[2]{\it Department of Radiology, Lausanne University Hospital and University of Lausanne, Lausanne, Switzerland.}
%\footnotetext[3]{\it LTS5, École Polytechnique Fédérale de Lausanne, Lausanne, Switzerland.}
%\footnotetext[4]{\it Bernard and Irene Schwartz Center for Biomedical Imaging, Department of Radiology, New York University School of Medicine, New York, NY, USA.}
%\footnotetext[5]{\it Center for Advanced Imaging Innovation and Research (CAI2R), Department of Radiology, New York University School of Medicine, New York, NY, USA.}
%\footnotetext[6]{\it Basel, Switzerland.}
%\footnotetext[7]{\it The Sackler Institute at the New York University School of Medicine, 550 First Avenue, New York, NY 10016, USA.\\ $^*$Correspondence; tom.hilbert@siemens-healthineers.com}

\section{Introduction}
Quantitative magnetic resonance (MR) measurements strive to estimate tissue-specific parameters with minimal experimental bias. Until recently, such methods have mostly focused on relatively simple spin evolutions for which analytic signal solutions can be derived. Early techniques to measure the relaxation time, for example, relied on a series of inversion-recovery measurements to estimate the longitudinal relaxation time ($T_1$) \cite{look1970time, pykett1983measurement} and on spin-echo measurements to estimate the transverse relaxation time ($T_2$)\cite{carr1954effects, meiboom1958modified}. Although such measurements can provide excellent results, they are generally too time-consuming to be used in routine clinical examinations.

For years, the search for faster methods has strived to achieve a balance between acquisition speed, model simplicity, accuracy, and precision\cite{velikina2013accelerating, zhang2015accelerating, block2009model}. One of the most widely used approaches in recent years is the combination of DESPOT1 and DESPOT2 techniques\cite{deoni2004determination}, combining four (or more) fast measurements to quantify both $T_1$ and $T_2$. Although these techniques are fast and SNR efficient, they are also sensitive to experimental imperfections\cite{sung2013transmit, hurley2012simultaneous} and magnetization transfer (MT) effects\cite{zhang2015does}.

The effect of MT on $T_1$ and $T_2$ quantification is especially strong in the brain, where it is significantly correlated with myelin content and axonal count\cite{schmierer2004magnetization}. Therefore, MT effects can also be repurposed as a biomarker for neurological diseases in which the myelination of the brain is altered, e.g. in multiple sclerosis\cite{vavasour1998comparison}. However, MT effects cannot be described by the basic Bloch equations, which are used for the signal description in most rapid quantitative MRI techniques. If these sequences are simulated using comprehensive models, a dependency of the model on additional experimental factors, such as properties of the RF pulses, becomes apparent. This dependency can influence the $T_1$ or $T_2$ estimation accuracy\cite{malik2018extended}. In theory, all these effects can be corrected for using information from separate measurements. However, each additional scan increases the acquisition time and adds complexity, such as the requirement for co-registration in the case of inter-scan motion. Moreover, each additional parameter increases the complexity of the analytical form of the signal model.

Recently, a new framework for quantitative MRI -- magnetic resonance fingerprinting (MRF) -- was proposed\cite{ma2013magnetic}. MRF moves away from comparatively simple steady state sequences and from straightforward analytic solutions. Instead, it combines more diverse sequence patterns, which produce transient states, with a numerical signal model that describes the corresponding spin dynamics. The additional degrees of freedom that become available with MRF enable faster imaging and provide the opportunity to deliberately encode multiple tissue properties and experimental conditions within a single measurement. MRF has sparked great interest in the research community because it may enable a transition from qualitative to quantitative MRI with short examination times, which would result in an extremely powerful tool for neuroscience and various clinical applications. The role of MT effects on MRF, however, has only begun to be investigated\cite{malik2018extended, hilbert2017mitigating}.

In this work, we explore the effect of MT on the quantification of $T_1$ and $T_2$ values based on MRF measurements. Specifically, a two-pool model is compared to a conventional one-pool model in phantom samples that are known to exhibit MT effects. The possibility to use this two-pool model for quantifying the fractional pool size of the semi-solid pool is studied in vivo. 

\section{Methods}
All numerical simulations, data analysis, and visualizations were performed using MATLAB 8.5 (The MathWorks Inc., Natick, MA, USA).

\subsection{Sequence Design}
To study the impact of MT, we used an MRF sequence design proposed by Cloos et al.\cite{cloos2018rapid} as a starting point. This baseline sequence will be referred to as ``Inversion Recovery FISP FLASH (IRFF)''. The IRFF sequence (see also Figure 1) consists of an adiabatic inversion pulse followed by four segments of RF pulse trains with flip angles up to 60° played out with constant repetition time TR of 7.5 ms. The first and second segments are steady state free precession RF trains (FISP-type). In the third and last segments, the transverse magnetization is spoiled using a quadratically increasing RF phase (FLASH-type, 50$^\circ$  phase increment). Gaps between the segments with a duration of 50 $\times$ TR allow spin ensembles to relax. In this work, the first and second gap were used to optionally play out 50 off resonant pulses (each with 7ms duration, 180$^\circ$ flip angle at 5kHz off-resonance frequency with Gaussian waveform). The hypothesis is that the additional pulses will only have an impact on the measured magnetization in the presence of a semisolid pool and, thus, may improve the encoding of MT effects within the fingerprint. In the following, we will refer to the sequence without MT pulses as IRFF and to the sequence with MT pulses as IRFF-MT.

\begin{figure}[h!]
  \includegraphics[width=\linewidth]{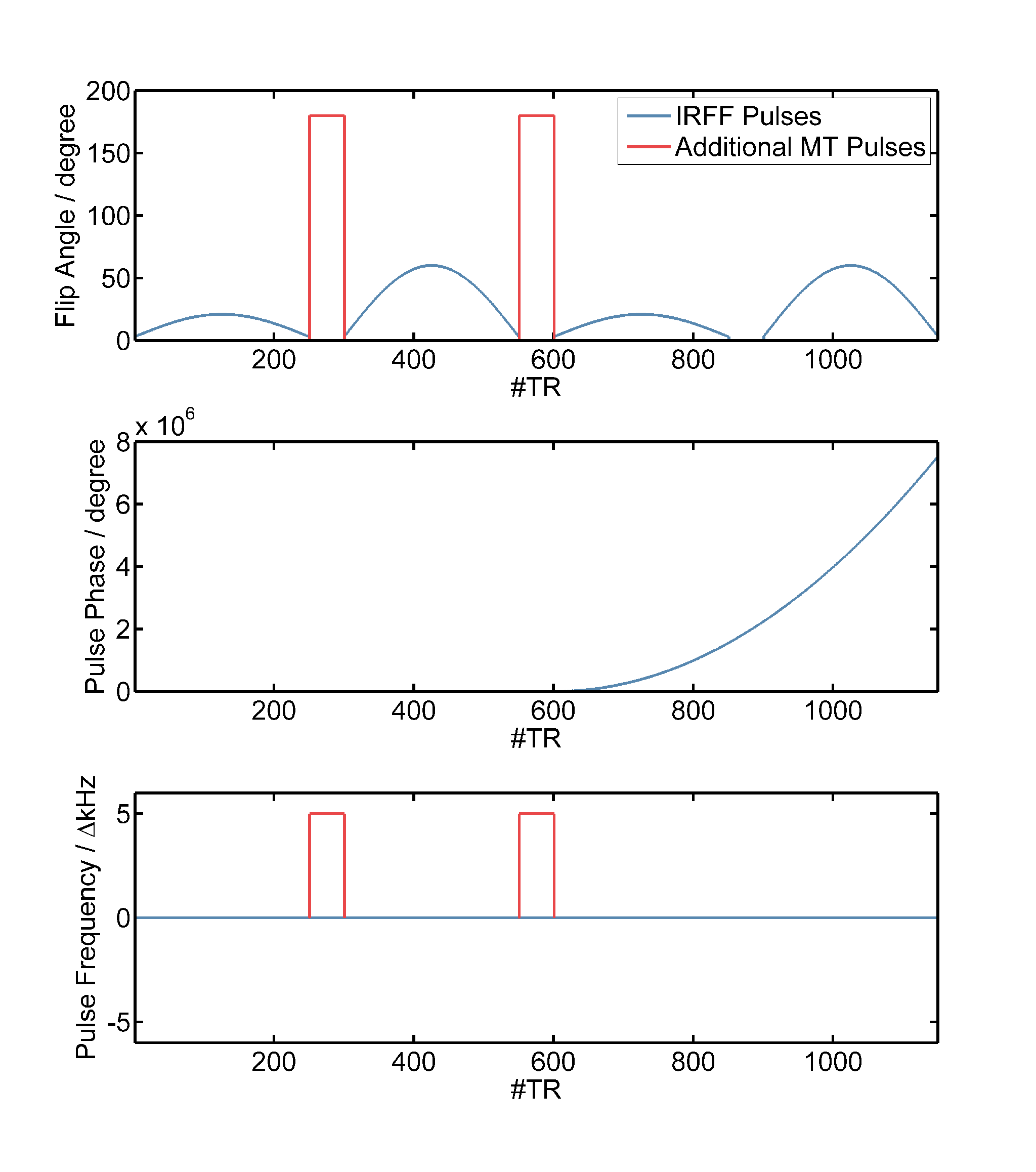}
  \caption{Sequence design used in this study with a train of flip angles (top) for each TR according to the IRFF sequence (blue) and optional MT pulses for the IRFF-MT design (red). The RF pulse train is separated into four segments (2x FLASH and 2x FISP) with gaps that allow for relaxation or to provide the space to play out the MT pulses. The phase of the pulses is shown in the middle with a quadratically increasing pulse phase for the last two FLASH segments. The off-resonance frequencies for the pulses are shown on the bottom.}
  \label{fig:seq}
\end{figure}

\subsection{Image Reconstruction}
Quantitative maps were reconstructed from the data using the reconstruction algorithm described by Cloos et al.\cite{cloos2018rapid}. However, a different Extended Phase Graph\cite{hennig2004calculation, weigel2015extended} algorithm was employed. Specifically, the EPG-X framework\cite{malik2018extended} was used to simulate fingerprints based on a single-pool or a two-pool model.

For the single-pool model, the simplest EPG-X model was used to simulate the MR signal based on the IRFF sequence design and a range of $T_1$, $T_2$, and $B_1^+$ values. To this end, the complex slice profile was discretized into 16 bins, and the signal was simulated for each bin. Subsequently, all fingerprints across the slice profile were summed, which yields the final fingerprint for the combination of quantitative values.

In order to account for MT effects, the two-pool model of the EPG-X framework was used. This model requires additional information about the deposited RF pulse power. To this end, the pulse power of each individual pulse (including the initial inversion pulse) in the IRFF sequence was calculated using the pulse duration, waveform, and flip angle (depending on $B_1^+$). Besides $T_1$, $T_2$, and $B_1^+$, the simulation with a two-pool model requires additional tissue parameters. Specifically, the relaxation parameters of the semi-solid pool $T_{1,ss}$, $T_{2,ss}$, the fractional semi-solid pool size F, and the exchange rate of magnetization from the free to the semi-solid pool k are required. This would result in four additional dimensions for the dictionary in order to address MT effects, which would yield large dictionaries with impractical reconstruction times. Furthermore, there may not be enough MT information encoded within the fingerprints to resolve these four additional parameters accurately without confounds. Therefore, several assumptions were made to model MT with only one additional dimension in the dictionary. First, it was assumed that the longitudinal relaxation times of the free and semi-solid pool are identical ($T_{1,ss}$ = $T_1$) \cite{gloor2008quantitative}. Furthermore, the transverse relaxation of the semi-solid pool, which only affects the shape of the frequency spectrum, was set to a fixed value according to literature\cite{gloor2008quantitative} ($T_{2,ss}$ = 12 $\mu$s). Similarly, the magnetization exchange rate was fixed to a value that is expected in white matter (WM) (k = 4.3 s$^{-1}$)\cite{gloor2008quantitative}. By introducing these assumptions, the fractional pool size $F$ remains as the only parameter to model MT. The systematic bias that is introduced by fixing the above model parameters was studied in more detail in Supporting Information S1.

Using the models described above, three dictionaries were created and used throughout this paper:
\begin{enumerate}
	\item {\bf Single-Pool IRRF Dictionary:} 190,527 entries (70 x $T_1$ ranging from 0.1 -- 4.3 s, 70 x $T_2$ ranging from 15 -- 430 ms and 41 x nominal $B_1^+$ ranging from 0.7 -- 1.3, corresponding to 2.9 GB of memory; entries with $T_2$ $\geq$ $T_1$ were excluded).
	\item {\bf Two-Pool IRRF Dictionary:} 3,048,432 entries (46.8 GB of memory) resulting from an additional 16 values in the $F$ dimension (ranging from 0 - 30\%, logarithmically spaced).
	\item {\bf Two-Pool IRRF-MT Dictionary:} This dictionary has the same number of entries as the two-pool IRRF dictionary above, but accounts for the off-resonance pulses in the IRRF-MT sequence.
\end{enumerate}

\subsection{In Vitro Experiments}
All in vitro experiments were performed using a whole-body 3-Tesla MRI system (MAGNETOM Prisma, Siemens Healthcare, Erlangen, Germany). A QED (Quality Electrodynamics, Mayville, OH, USA) 15-channel TX/RX knee coil was used for excitation and reception

Both prototype sequence configurations, IRFF and IRFF-MT, were acquired without phase-encoding gradients to directly sample the MRF signal (i.e., the fingerprints) from two phantoms. The first phantom was a tube filled with water that was doped with Manganese (II) Chloride Tetrahydrate (Cl2Mn4H20, Sigma-Aldrich, St Louis, MO, USA) to serve as a sample without MT. For the second phantom, cross-linked bovine serum albumin (xl-BSA) with a final concentration of 20\% (w/w) was prepared using PBS and Glutaraldehyde\cite{Kroh2018IMMOBILISE} to serve as a sample that exhibits MT. The hypothesis is that fingerprints from IRFF and IRFF-MT should be identical in both water and xl-BSA samples if there were no MT effect. Conversely, if the fingerprints are affected by MT, the signal should decrease in the second and third segment due to the MT pulses employed.

MR signals were plotted for comparison between both samples with different sequence types, respectively. Furthermore, all three dictionaries were used to match the signals acquired from both water and xl-BSA samples. The measured signals and the corresponding best match were plotted for each dictionary (all normalized by the $l_2$ norm), and the corresponding quantitative parameters ($T_1$, $T_2$, ${B_{1}}^{+}$, $F$) were compared. Furthermore, the normalized root-mean-square error (NRMSE) between the measured and best matched simulated signal was calculated as follows:
\begin{equation}
NRMSE = \frac{|| s-f ||_2}{|| s ||_2},
\end{equation}
where $s$ denotes a vector of measured signal intensities and $f$ a vector of simulated signal intensities.

\subsection{In Vivo Experiments}
All in vivo experiments were performed using a whole-body 3-Tesla MRI system (MAGNETOM Skyra, Siemens Healthcare, Erlangen, Germany). The built-in birdcage body coil was used for excitation, and a commercially available 64-channel head/neck coil was used for reception.

The IRRF and IRFF-MT prototype sequences were used to acquire datasets from five healthy volunteers (two female, age range 21-33 years) after written informed consent was obtained prior to the examination. A single axial slice through the brain was imaged using a matrix size of 256x256, 256 mm FOV, 4 mm slice thickness, TR of 7.5 ms, and 10 radial spokes per time point of the fingerprint series, resulting in TA = 3:10 min scan time for each dataset. The study was approved by our institutional review board. 

All datasets were reconstructed using the three dictionaries as described above. The IRFF dataset was reconstructed with both the single-pool and the two-pool model, whereas the IRFF-MT dataset was reconstructed with the two-pool model only.

For comparison, a reference $T_1$ map was acquired using an MP2RAGE sequence, and reference $T_2$ values were acquired using a Carr-Purcell-Meiboom-Gill (CPMG)\cite{meiboom1958modified} sequence and a dictionary matching that accounts for stimulated echoes\cite{lebel2010transverse}. Reference MT values were obtained from an MTR acquisition (two FLASH acquisitions), and reference B1 values were obtained with two saturation-prepared FLASH acquisitions. It should be noted that these techniques are not gold-standard methods since they are also affected by model assumptions, such as considering a single compartment with no MT effects, and no diffusion effects. However, a comparison to existing methods should help to put the results obtained here into context.

In order to compare the quantitative values of different brain regions, a 3D MPRAGE sequence was used as input for tissue segmentation with the MorphoBox prototype\cite{schmitter2015evaluation}. Detailed sequence parameters are provided in Table 1. 

\begin{figure}[b]
  \includegraphics[width=\linewidth]{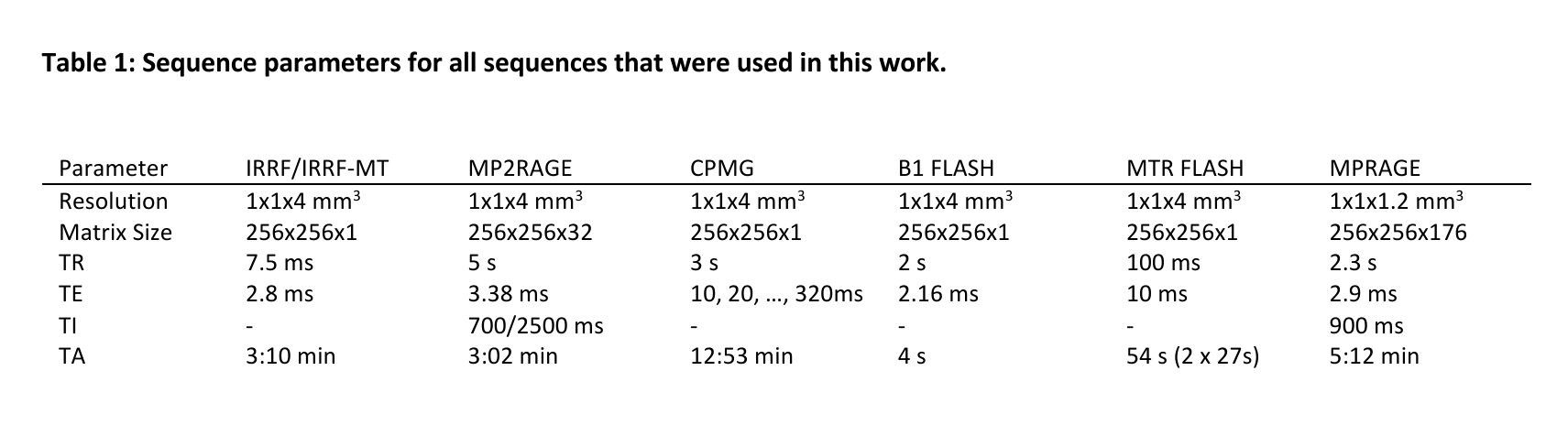}
  \label{fig:tab1}
\end{figure}

After the acquisition, the maps obtained from the IRFF/IRFF-MT sequences and different dictionaries as well as from the reference sequences were compared. To this end, the MPRAGE image was registered to the images of all sequences, and the same transformation was applied to the label map of the MorphoBox segmentation to have the same segmentation in the native spaces of all quantitative maps. The median values of $T_1$, $T_2$, and F/MTR within eight bilateral regions (frontal WM, parietal WM, frontal GM, parietal GM, Corpus Callosum, Thalamus, Caudate, Putamen) were extracted for each subject. The mean and standard deviation across subjects from the different regions were compared between the different dictionaries and reference methods.

\section{Results}
\begin{figure}[b]
  \includegraphics[width=\linewidth]{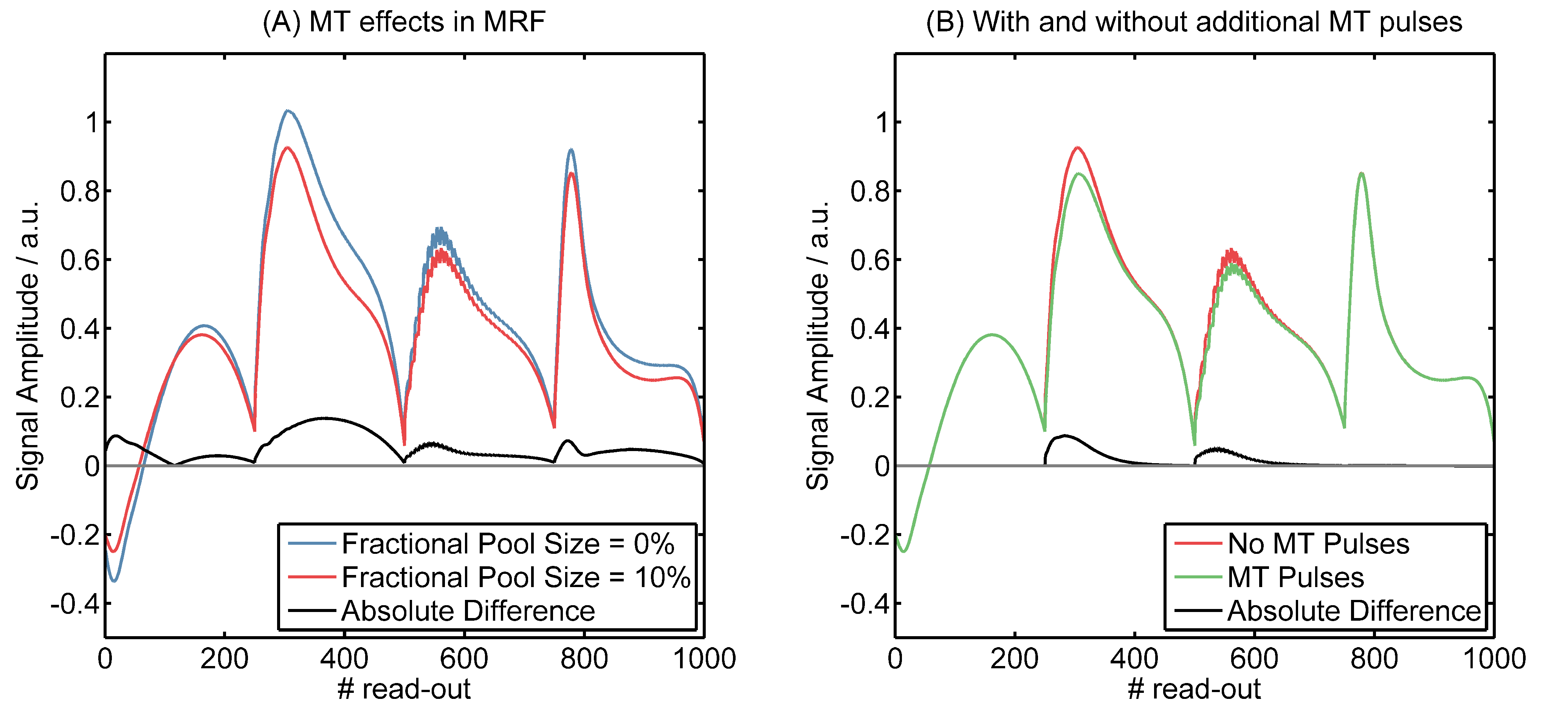}
  \caption{(A) Example dictionary entries that show the difference in fingerprints due to MT (no MT versus a fractional pool size of 10~\%) for a typical WM voxel (T1 = 800 ms, T2 = 60 ms, $B_1^+$ = 1) without MT pulses. (B) Differences in fingerprints ($T_1$ = 800 ms, $T_2$ = 60 ms, $B_1^+$ = 1) that experience MT (fractional pool size = 10~\%) due to adding off-resonance MT pulses in the first and second gap in the sequence design.}
  \label{fig:sim}
\end{figure}
\subsection{Simulated Fingerprints \& Dictionaries}

The creation of the dictionaries took approximately 10 h for the single-pool and 7 days for the two-pool model using 24 cores on an Intel® Xeon® Gold 6126 CPU at 2.60 GHz. The dictionary matching required approximately 5:49 min for the single-pool and 1 h 42 min for the two-pool model using a single CPU core (same specification as above).

Example dictionary entries are shown in Figure 2. Figure 2a demonstrates the impact of MT on the fingerprint by comparing dictionary entries with the same relaxation parameters but different fractional pool size ($F$ = 0 \% versus $F$ = 10 \%). The largest differences were observed at the beginning after the high-power adiabatic inversion pulse and in the second and last segments where higher flip angles were used. Figure 2b compares dictionary entries from the IRFF and IRFF-MT sequence designs at a fractional pool size of 10 \% (an expected value in WM). The first and last segments were almost identical. However, in the second and third segment, lower signal intensity was observed in the IRFF-MT sequence due to the MT effect of the semi-solid pool, which is amplified by the off-resonant MT pulses during the gaps between the segments.

\subsection{In Vitro Experiments}
The MR signals from doped water and xl-BSA that were acquired with IRFF and IRFF-MT are shown in Figure 3. In the water sample, where no MT effects are expected, the additional off-resonance MT pulses did not show an effect. The measured signals from IRFF and IRFF-MT were almost identical. In contrast, in the xl-BSA sample, the signal in the second and third segments was lower for the IRFF-MT sequence compared to the IRFF sequence, similar to the comparison of dictionary entries in Figure 2b.
\begin{figure}[h!]
  \includegraphics[width=\linewidth]{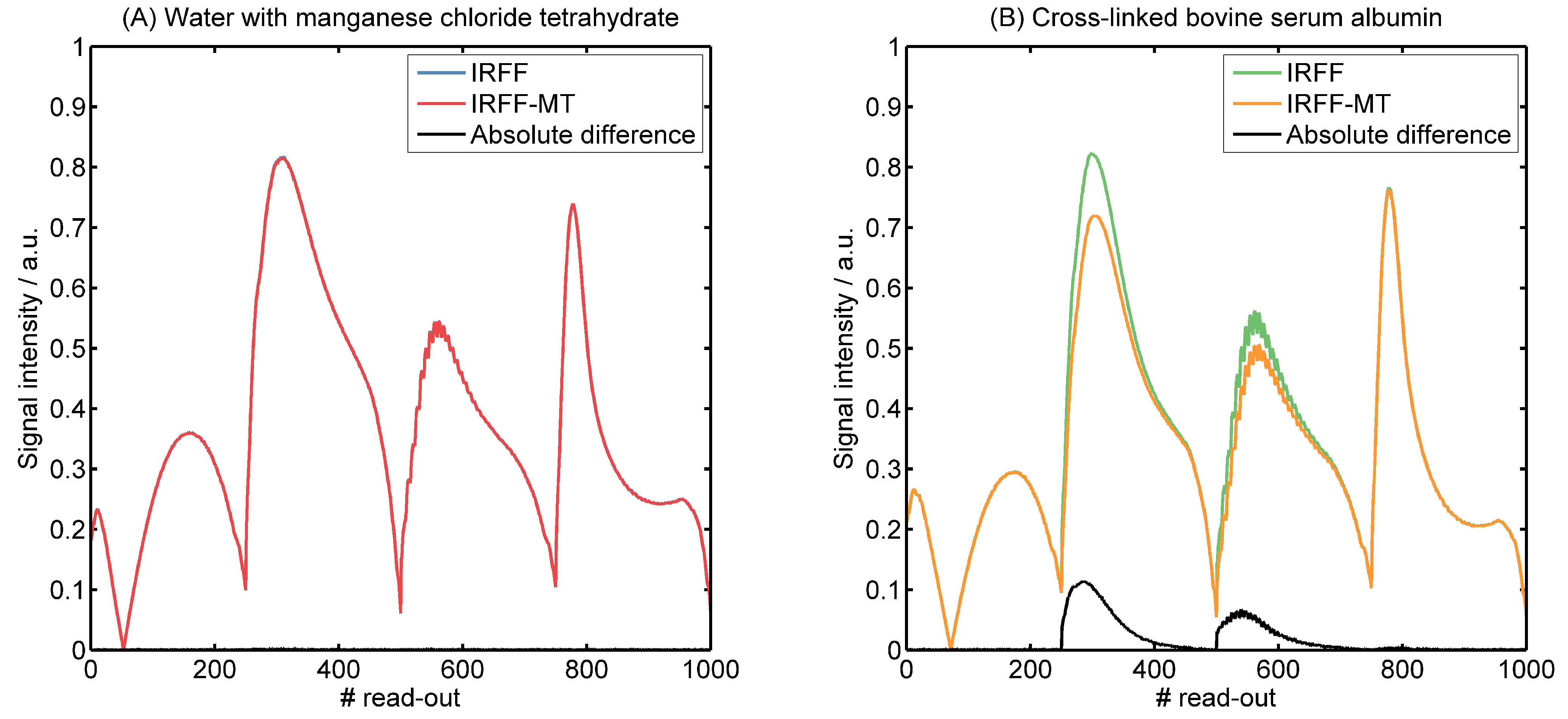}
  \caption{Four different acquired MR signals (fingerprints) from the water (left, no MT effect) and xl-BSA (right, MT effect) samples using the conventional fingerprinting sequence design (IRFF) and one with off resonant MT pulses in the gaps of the sequence (IRFF-MT).}
  \label{fig:meas}
\end{figure}
\begin{figure}[h!]
  \includegraphics[width=\linewidth]{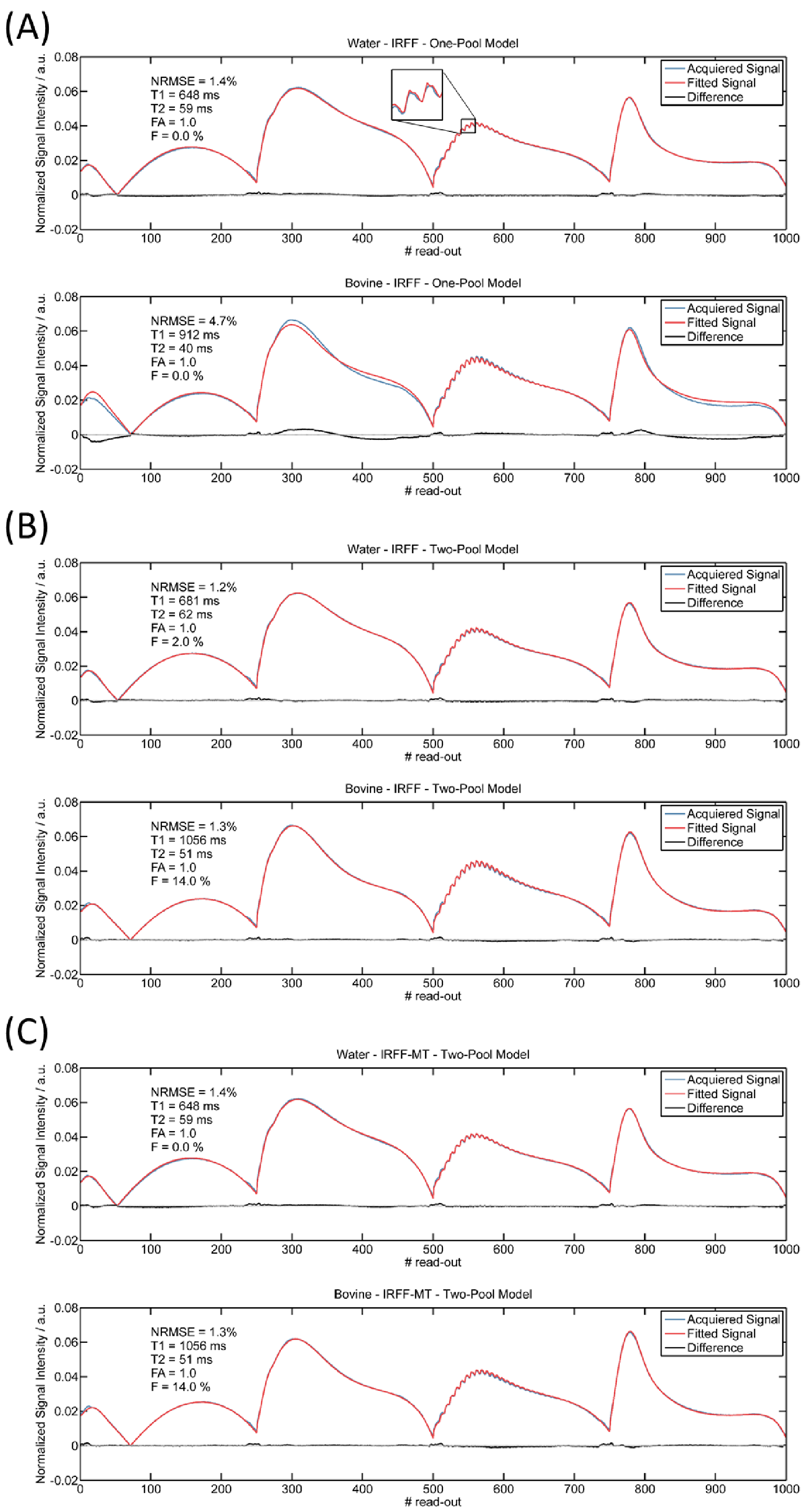}
  \caption{Measured MR signals from doped water (no MT effect) and xl-BSA (MT effect) and corresponding best matching dictionary entries from the different dictionaries: (A) single-pool model, (B) two-pool model, (C) two-pool model with MT pulses (IRFF-MT).}
  \label{fig:meas2}
\end{figure}
Figure 4a shows fingerprints that were acquired with the IRFF sequence and their corresponding best matches from the single-pool dictionary. In the water sample, the best matching dictionary entry corresponded well to the acquired data (NMRSE = 1.4~\%), and even the oscillations in the third segment agreed well. However, in the xl-BSA sample, the best matching single-pool dictionary entry showed large differences at the beginning (after the inversion pulse), and in the segments with larger flip angles (second and fourth segments), thus, resulted in an overall larger NRMSE of 4.7~\%. A much lower NRMSE of 1.3~\% can be achieved if a two-pool model is matched to the xl-BSA sample (Figure 4b, bottom). Notably, in comparison to the single-pool model, the two-pool model resulted in longer $T_1$ (1056 ms versus 912 ms) and $T_2$ (51 ms versus 40 ms) estimates. The best matching two-pool entry for the water sample showed an even lower NRMSE than the single-pool model (1.2~\% versus 1.4~\%). However, also a fractional pool size $F$ = 2~\% was observed. This means that a better matching dictionary entry was found with MT than without MT. Assuming that there should be no MT in water, this could indicate that MT effects cannot be well separated from relaxation effects in the signal of the IRFF sequence. Figure 4c shows fingerprints and best matching signals of the two-pool model from the IRFF-MT sequence. After the introduction of the off-resonance pulses in the sequence design, the model describes the fingerprint of the xl-BSA sample well and resulted in the same sample properties ($T_1$ = 1056 ms, $T_2$ = 51 ms, $B_1^+$ = 0.98, $F$ = 14 \%). In the water sample, however, the introduction of the MT pulses resulted in a better differentiation between relaxation and MT effects, yielding a 0 \% fractional pool size and the same sample properties as for the single-pool model ($T_1$ = 648 ms, $T_2$ = 29 ms, $B_1^+$ = 0.98).

\subsection{In Vivo Experiments}

Example quantitative maps obtained from one volunteer using the three different dictionaries are shown in Figure 5. Regardless of the applied dictionary, all PD maps were affected by receive-field inhomogeneity and, unlike other parameter maps, should not be considered fully quantitative. Nevertheless, contrast between white and gray matter was observed in the PD maps reconstructed from the single-pool model, with lower PD in gray matter in comparison to white matter. Furthermore, in both $T_1$ and $T_2$ maps, lower relaxation values were observed when a single-pool model was used. Since this effect is stronger in white matter, the contrast between white and gray matter was reduced when using a two-pool model. Of note, the increase in $T_1$ and $T_2$ when accounting for MT was observed in vitro as well (see Figure 4a-b). The fractional pool size of the semi-solid pool (F) corresponds to a contrast that resembles MT maps in literature\cite{gloor2008quantitative, weiskopf2013quantitative, dortch2011quantitative}  with the highest fractional pool size in white matter, lower fractional pool size in gray matter, and no MT effect in CSF. When comparing IRFF versus IRFF-MT for the two-pool model, the latter, which employs MT pulses, resulted in lower fractional pool size. This may be linked to an overestimation of $F$ if MT effects are not sufficiently encoded in the IRFF fingerprint, as also demonstrated in vitro (see Figure 4b-c). The obtained $B_1^+$ maps showed high flip angles in the center of the FOV with a smooth transition to lower flip angles in the periphery of the FOV. The $B_1+$ field maps obtained from the different dictionaries appeared similar. 

\begin{figure*}[h!]
  \includegraphics[width=\linewidth]{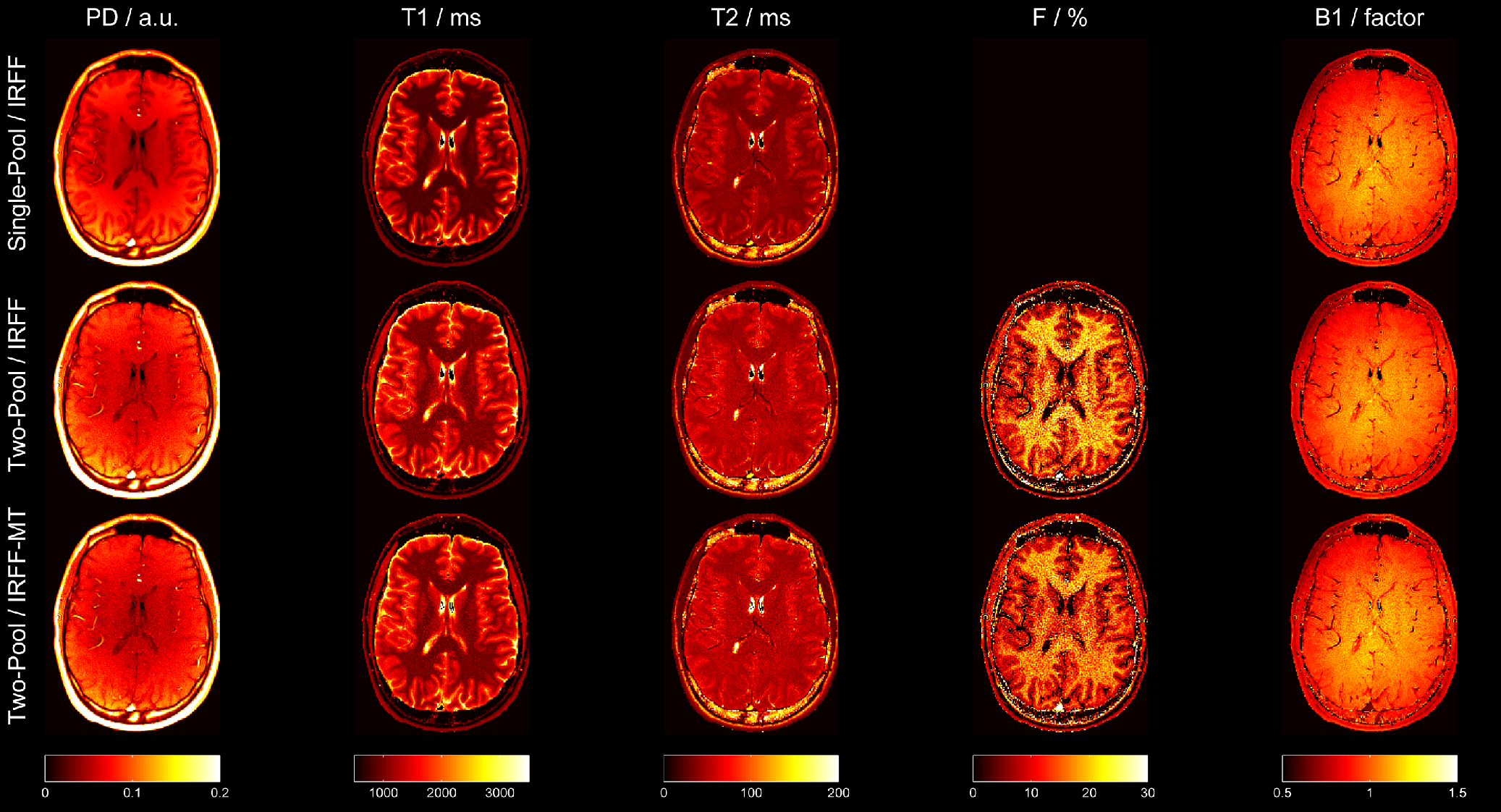}
  \caption{Example quantitative maps obtained from one volunteer using a single-pool model (top), a two-pool model (middle) or a two-pool model with MT pulses in the sequence (IRFF-MT, bottom).\newline ~ \newline ~}
  \label{fig:inv}
\end{figure*}

Quantitative values obtained from all subjects using the three different dictionaries and reference sequences are shown in Table 2. In WM, the single-pool model estimated similar $T_1$ values compared to the reference MP2RAGE sequence (ranging from 759 - 860 ms). In contrast, both two-pool models yielded higher $T_1$ values (ranging from 996 - 1087 ms). $T_1$ values of GM were between 1334 ms and 1377 ms for the single-pool model and the MP2RAGE. Again, higher $T_1$ values were found for the two-pool models (ranging from 1515 ms - 1530 ms).

\begin{figure*}[h!]
  \includegraphics[width=\linewidth]{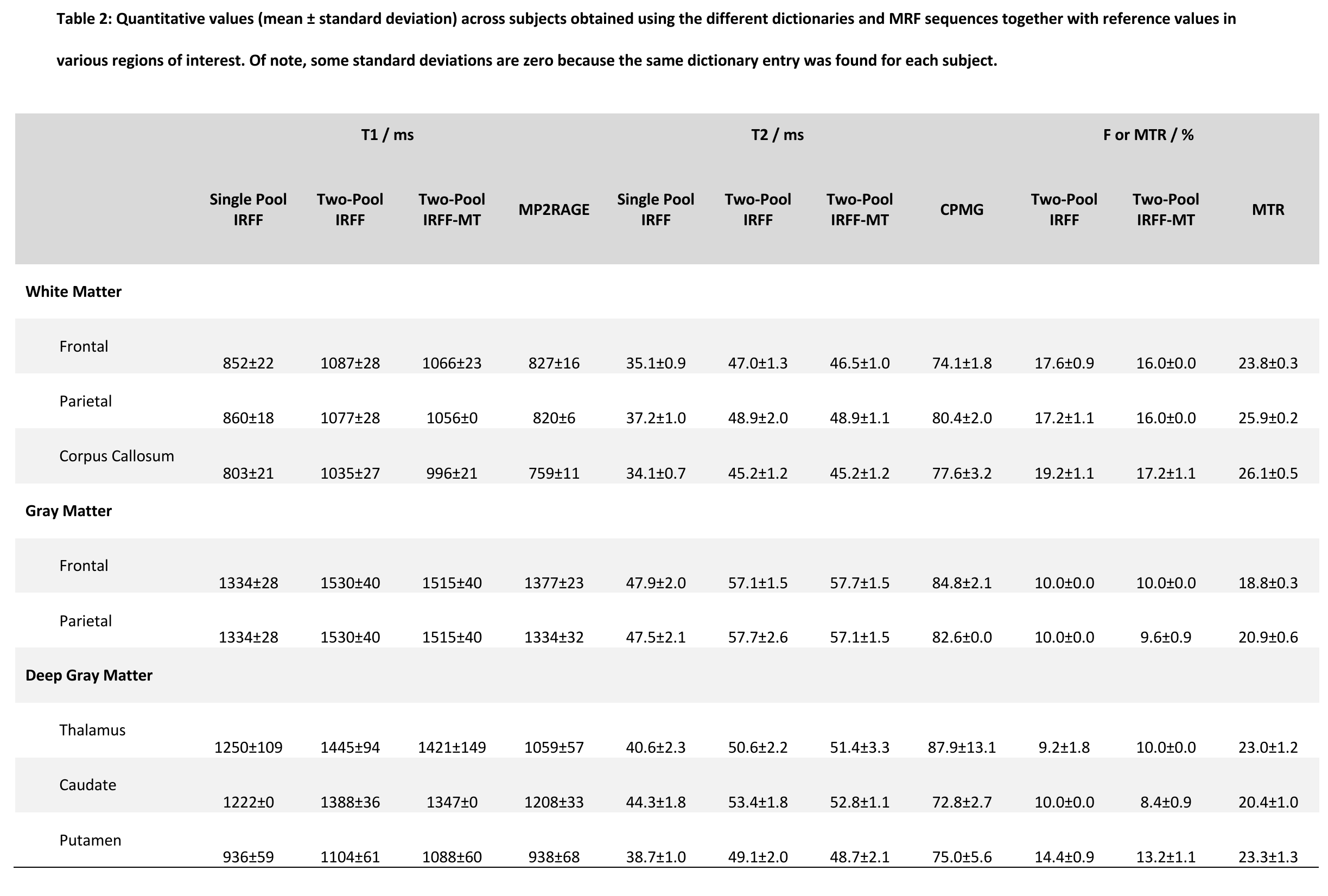}
  \label{fig:tab2}
\end{figure*}

For $T_2$ values, the single-pool model provided the lowest values (ranging 34.1 ms - 47.9 ms), which increased once MT was accounted for (ranging from 45.2 - 57.7 ms). The $T_2$ values obtained with the CPMG sequence were always higher (ranging from 74.1 - 87.9 ms). 

The measured fractional pool size in WM was always lower when MT pulses were employed in the sequence (17~\% for IRFF and 16~\% for IRFF-MT). The fractional pool size was lower in gray matter ($\approx$10 \%). The MTR, which is the ratio between two images, with and without MT pulses, was higher than the estimated fractional pool size, suggesting that MTR is not a reliably quantitative measure. However, the fractional pool size correlated well with MTR, with slightly better Pearson correlation for IRFF-MT (0.84 and 0.87 for IRFF and IRFF-MT respectively).

\section{Discussion}
Our results provide evidence of the possible influence of MT when using MRF to quantify $T_1$ and $T_2$ values in the brain. Based on these results, we propose using a two-pool model to make the estimation of relaxation parameters less susceptible to eventual MT effects and to quantify the fractional pool size.

When using a two-pool model, a better match between measured and simulated fingerprints can be achieved, as demonstrated for in vitro MR signals. However, without modification of the sequence design, MT effects cannot be fully separated from relaxation effects in the presence of undersampling artifacts and noise. We included off-resonance MT pulses to improve the encoding of MT in the fingerprints, which resulted in an overall better match. Of note, the match for signals without MT is still improved in the presence of MT pulses since it is less likely that relaxation effects can be mistaken for MT effects.

The two-pool model used in this study was simplified by fixing two of the model parameters to literature values. It was assumed that brain tissue does not have more than two pools, that it involves no inhomogeneous MT\cite{varma2015magnetization} and it is well represented by a single compartment\cite{mackay1994vivo,harrison1995magnetization}. Moreover, physiological effects, such as diffusion and perfusion, were neglected. The proposed model is, thus, still an approximation of the actual microstructural and biochemical environment. Ideally, the simulation of the fingerprints should use a complete model of the tissue microstructure. However, apart from the computation time needed to calculate such dictionary, the matching robustness would likely suffer because subtle microstructural effects may be poorly encoded in the fingerprint. For example, although MT is better reflected in the fingerprint by including MT pulses in the sequence, the difference is still small, which resulted in unstable matching, as shown by the relatively noisy $F$ maps. 

\newpage
~
\newpage
~
\newpage
In a wider context, one open research question is, therefore, how much detail is required to model tissue microstructure and how well these effects can be encoded within a fingerprint. Besides employing MT off-resonance pulses, there may be other acquisition techniques that will improve the encoding of MT. In an early stage of this work, we attempted to encode MT by varying the pulse duration\cite{hilbert2017mitigating}. However, the signal behavior was rather difficult to model, presumably due to relaxation during the RF pulse. For instance, the fingerprints changed even in the water sample when the pulse duration was varied (see Supporting Information S2). 

Alternatively, one could also attempt to minimize MT bias effects by desensitizing the sequence to MT. For example, a longer TR would allow the semi-solid pool to recover more longitudinal magnetization between excitation pulses. However, this would lead to less efficient data acquisition. Longer pulse durations with a narrower saturation profile in the frequency domain could also be considered to minimize the effect on the semi-solid pool. However, longer pulse duration would also require longer TR, leading again to a less efficient data acquisition. The sequence design, i.e., flip angles and spoilers, has an impact on the MT effect as well. Here, the experiments used the IRFF sequence design. Other sequence designs may be more or less sensitive to MT and may result in a different bias in terms of the effect size and direction (overestimation vs. underestimation). This should be further investigated.

The reference methods (MP2RAGE, CPMG) also use simplified models to accommodate feasible acquisition times and to condition fitting procedures, as is typically the case for quantitative mapping approaches. Therefore, these methods may also suffer from a systematic bias due to MT or other contrast mechanisms. This may explain the relatively large discrepancy of quantitative values across the literature\cite{stikov2015accuracy}.

Future work should also include improved phantom design to better represent the true microstructure of tissue and to incorporate multiple pools or compartments. The in vitro experiments described here used an in-house preparation of xl-BSA, which may vary between institutions. Therefore, extensive validation of the effect of MT on quantitative parameter estimation is challenging, as standardized phantoms are not available.

For simplicity, all experiments were performed using single-slice acquisitions. The acquisition of multiple slices, interleaved or sequential, will introduce saturation of the semi-solid pool that needs to be accounted for in the spin history: the on-resonant pulse from a slice causes off-resonance saturation of the semi-solid pool in other slices, depending on the relative slice distance and the slice-selection gradient\cite{dixon1990incidental, santyr1993magnetization}. The model proposed here can be easily extended to interleaved multi-slice acquisition by applying additional shifted saturation profiles, according to the slice order. This modification will, however, restrict the slice parameters of the acquisition protocol to the trained dictionary.

In general, the MT effect and its impact on the accuracy of the relaxation estimation should be further studied in the context of quantitative imaging. Also, the sequence design proposed here should be further investigated to improve SNR efficiency and to explore the possibility of a more complete quantification of the MT parameters, e.g. independently estimating the exchange rate and fractional pool size. Quantitative MT would be of high interest for many clinical applications. For example, in the brain, as the myelin sheaths surrounding the axons mainly consist of macromolecules. Therefore, it may potentially be applicable as an early marker for demyelination\cite{schmierer2004magnetization,vavasour1998comparison}.

A major limitation of the proposed method is that the current acquisition time is 3:10 min for a single slice, which corresponds to a 1 h 35 min protocol for whole-brain coverage (assuming 30 slices). However, with advanced reconstruction techniques\cite{pierre2016multiscale,doneva2017matrix}, k-space trajectories\cite{ye2017simultaneous}, and 3D acquisitions\cite{liao20173d,ma2018fast}, it may be possible to achieve clinically acceptable acquisition times. The long reconstruction time is another limitation of the proposed method. However, different methods have been proposed recently to drastically reduce reconstruction times by using either non-linear kernels\cite{nataraj2018dictionary} or neural networks\cite{cohen2018mr}, which will be investigated in future work. 

\section{Conclusion}
Our work demonstrates that MRF relaxation-parameter estimation can be influenced by MT effects. To alleviate the impact of MT, we evaluated a two-pool model to match the data instead of a conventional single-pool model. In addition, we modified the original IRFF sequence to apply off-resonance MT pulses in the RF train, in order to better differentiate between MT and relaxation effects in the fingerprint. This work shows that different $T_1$ and $T_2$ values are obtained when accounting for MT effects and, furthermore, that fractional pool size ($F$) maps can be estimated along with other parameters in an MRF context.

\section*{Acknowledgements}
The authors thank Steffen Goerke from the German Cancer Research Center (DKFZ) for providing a protocol for preparing the high concentration cross-linked BSA.

This work was supported in part by NIH R21 EB020096, NIH R01 AR070297, and NIH R01 EB026456, and was performed under the rubric of the Center for Advanced Imaging Innovation and Research (CAI2R, www.cai2r.net), a NIBIB Biomedical Technology Resource Center (NIH P41 EB017183).

\bibliography{mtbib} 
\bibliographystyle{ieeetr}

\newpage
\onecolumn
\section*{\centering \bf Supporting Information S1}
{\noindent
{\bf Introduction: }In the main manuscript, two parameters ($T_{2,SS}$ and k) in the EPG-X simulations that govern magnetization transfer were held constant in order to reduce the dimensionality of the dictionary ($T_{2,ss}$ = 12 $\mu$s, k = 4.3 s$^{-1}$). Fixing these parameters will result in errors, since these parameters have a direct impact on the strength of the MT effect. The goal of this supporting information is to use simulations to evaluate the size of the associated errors, and to determine which estimates ($T_1$, $T_2$, $B_1^+$ or $F$) are affected the most.

{\noindent \bf Methods: }A synthetic dataset of fingerprints was generated for an image size of 128 x 128, where each voxel represents a different combination of ($T_{2,SS}$ and $k$). The remaining parameters were fixed ($T_1$ = 800 ms, $T_2$ = 60 ms, $B_1^+$ = 1, and $F$ = 10 ~\%). For all simulations, the two-pool model of the IRFF-MT sequence was used. Afterwards, the dictionary with fixed parameters (same as in the in vivo experiments) was used to estimate $T_1$, $T_2$, $B_1^+$, and $F$. The absolute error compared to the ground truth was calculated.

{\noindent \bf Results:} The error maps are shown in the figure below. A cross indicates the location corresponding to the correct assumed parameters. The $T_1$ error ranges from -50 to 27 ms at the edges of the tested range where mostly the extremes of $T_{2,SS}$ result in a bias of $T_1$ values. On the other hand, the error in $T_2$ (ranging from -7 to 5 ms) is more complex and mostly depends on which dictionary entry fits best with regards to $T_1$, $B_1^+$ and $F$. Notably, a small error in $T_1$ and $T_2$ is observed even when the assumed values of $T_{2,SS}$ and k are correct (at the cross in the figure) because of the discrete step size in the dictionary, i.e. there is no entry that exactly matches values of $T_1$ = 800 ms and $T_2$ = 60 ms. The estimation of $B_1^+$ is only affected by variations in $T_{2,SS}$ resulting in error ranging from -0.03 to 0.02 in nominal flip angle. The estimation of the fractional pool size $F$ is relatively robust to the assumptions, with errors between -4~\% and 2~\% with respect to the gold standard.

{\noindent \bf Discussion and Conclusion:} The error caused by assuming fixed values in the dictionary results in moderate biases of all estimated parameters. The estimation error is highly interdependent between the parameters due to the discrete solution space of the dictionary. In general, within the tested range $T_{2,SS}$ seems to have a larger impact on estimation errors in comparison to the exchange rate $k$. These errors should be considered when interpreting the quantitative results obtained with the proposed method.
}
\begin{figure}[h!]
  \includegraphics[width=\linewidth]{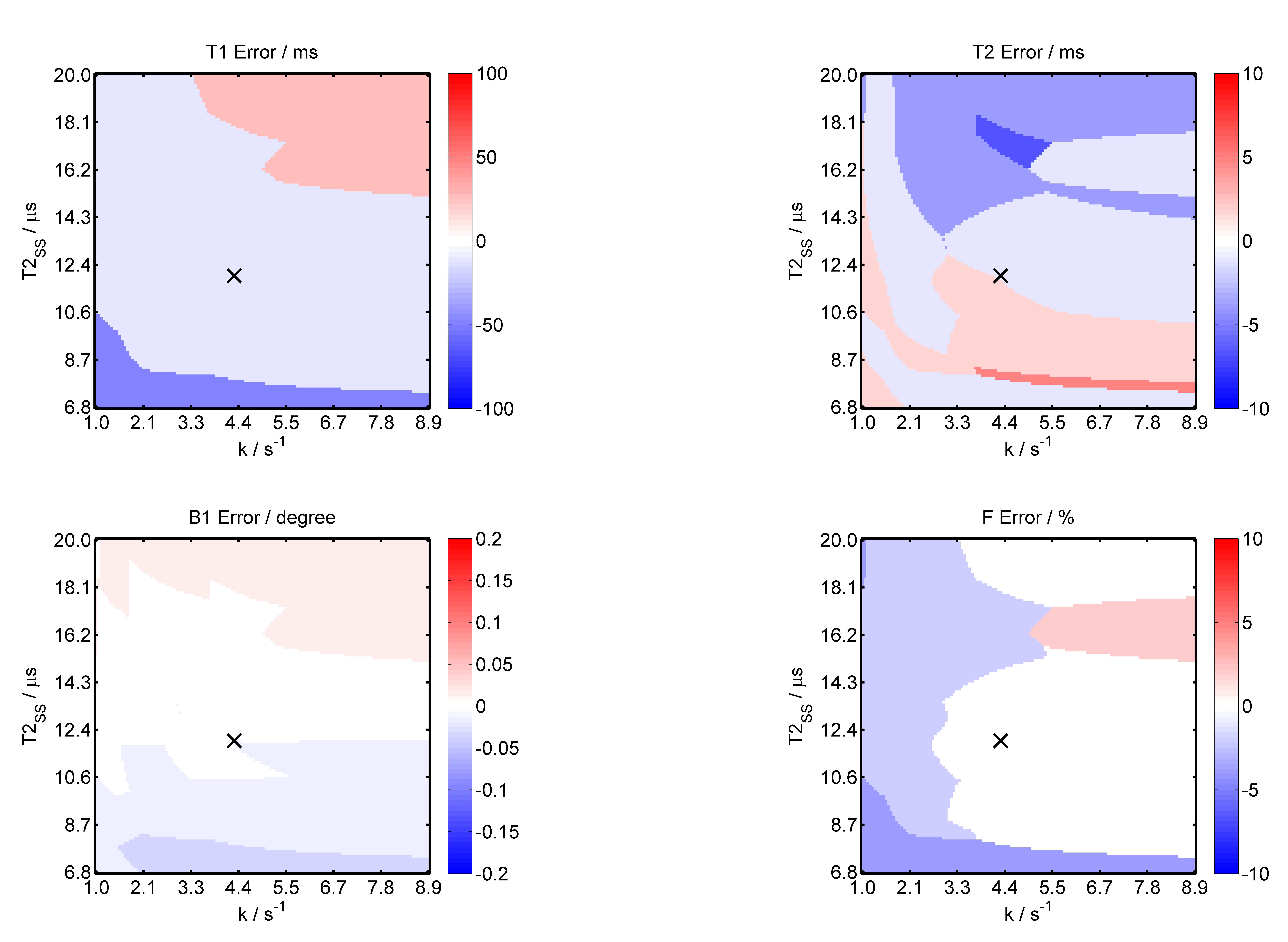}
    {Absolute estimation errors in $T_1$, $T_2$, $B_1^+$ and fractional pool size $F$ if the assumption of the fixed $T_2$ of the semisolid pool ($T_{2,SS}$) and exchange rate ($k$) is not met. The cross indicates the location where the assumption is true.}
  \label{fig:S1}
\end{figure}

\newpage

\section*{\centering \bf Supporting Information S2}
{\noindent
{\bf Introduction:} To improve the encoding of magnetization transfer (MT) in the fingerprint, it was proposed to introduce off resonance MT pulses in the first two gaps of the sequence. Alternatively, the MT effect can be varied by changing the applied pulse power during the acquisition of the fingerprint. This was tested in an early version of this work$^1$ by varying the pulse duration. In this supporting information we provide some of the early results that demonstrated that varying the pulse power may be confounded by other effects and may not be ideal for encoding MT in the fingerprint.

{\noindent \bf Methods:} The same IRRF sequence design was used to acquire MR signals from two samples (just as in the main manuscript: doped water, and xl-BSA). For both samples, the MR signal was acquired twice, each time with a different pulse duration (2, and 0.5 ms). The assumption is that the MR signals acquired with different pulse durations in the water are supposed to be identical since there is no MT expected. However, a difference in NMR signals is expected for the xl-BSA sample, since the different pulse durations (and associated pulse powers) will result in different MT effects.

{\noindent \bf Results:} For the xl-BSA sample, changing the pulse duration results in larger differences throughout the fingerprint but especially in segments with larger flip angles (see figure B below). The shorter pulse duration (i.e. larger pulse power) results in a lower signal intensity. However, the water sample also shows differences between MR signals with different pulse duration (see figure A below). These differences are smaller in comparison to the xl-BSA sample, but the shorter pulse duration resulted in higher signal intensities in water.

{\noindent \bf Discussion and Conclusion:} The differences that were observed in the water sample may be caused by relaxation during the pulse. These effects may be confounding since they will overlap with the MT effect that the method is intended to encode. Currently, the simulation used to create the dictionary assumes an instantaneous application of the pulse. Therefore, no differences can be observed when simulating the signals with different pulse durations, although we can observe effects experimentally.
In conclusion, if it is desired to measure MT by varying pulse durations, it will also be necessary to simulate potential confounding effects such as relaxation during the pulse application.
}

\begin{figure}[h!]
  \includegraphics[width=\linewidth]{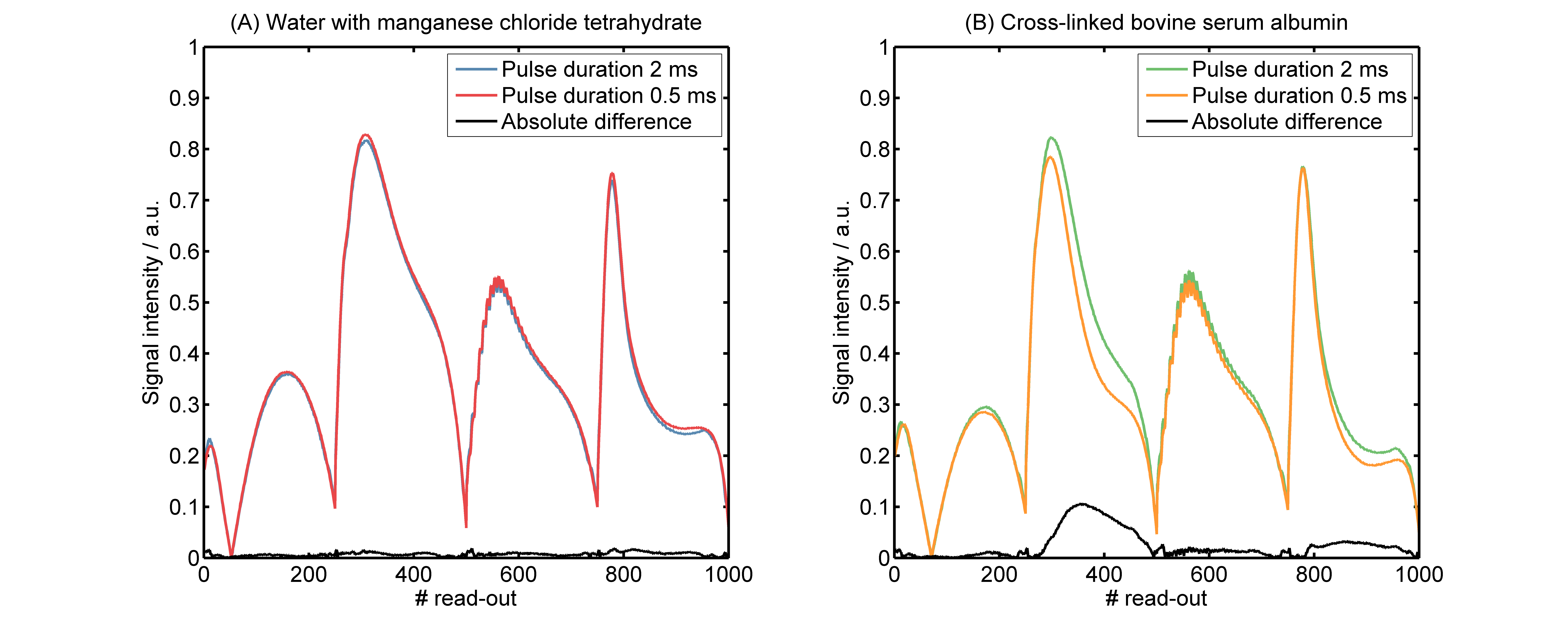}
    {Acquired MR signals (fingerprints) from the water and xl-BSA samples using the IRFF sequence with two different pulse durations of 2 ms and 0.5 ms.}
  \label{fig:S2}
\end{figure}

[1] Hilbert T, Kober T, Zhao T, et al. Mitigating the Effect of Magnetization Transfer in Magnetic Resonance Fingerprinting. In: International Society for Magnetic Resonance in Medicine. ; 2017:0074.

\end{document}